%% file: main.tex
\documentclass[AMA,STIX1COL]{WileyNJD-v2}
\usepackage{ulem}
\usepackage{stix}
\usepackage{amsmath}
\usepackage{threeparttable}
\usepackage{bbm}
\numberwithin{equation}{section}
\newcommand\independent{\protect\mathpalette{\protect\independenT}{\perp}}
\def\independenT#1#2{\mathrel{\rlap{$#1#2$}\mkern2mu{#1#2}}}
\newcommand{\distas}[1]{\mathbin{\overset{#1}{\kern\z@\sim}}}%
\newsavebox{\mybox}\newsavebox{\mysim}
\newcommand{\distras}[1]{%
  \savebox{\mybox}{\hbox{\kern3pt$\scriptstyle#1$\kern3pt}}%
  \savebox{\mysim}{\hbox{$\sim$}}%
  \mathbin{\overset{#1}{\kern\z@\resizebox{\wd\mybox}{\ht\mysim}{$\sim$}}}%
}

\articletype{Article Type}%
\usepackage{setspace}
\usepackage{pdflscape} 
\received{}
\revised{}
\accepted{}
        
\begin{document}

\title {A Hierarchical Meta-Analysis for Settings Involving Multiple Outcomes across Multiple Cohorts 
}


\author[1]{Tugba Akkaya Hocagil*}
\author[2,3]{Louise M. Ryan*}
\author[1]{Richard J. Cook}
\author[4]{Gale A. Richardson}
\author[4]{Nancy L. Day}
\author[5]{Claire D. Coles}
\author[6,7]{Heather Carmichael Olson}
\author[8]{Sandra W. Jacobson}
\author[8]{Joseph L. Jacobson}


\address[1]{\orgdiv{Department of Statistics and Actuarial Science}, \orgname{University of Waterloo}, \orgaddress{\state{ON}, \country{Canada}}}
\address[2]{\orgdiv{School of Mathematical and Physical Sciences}, \orgname{University Of Technology Sydney}, \orgaddress{\state{NSW}, \country{Australia}}}
\address[3]{\orgdiv{Department of Biostatistics}, \orgname{Harvard University}, \orgaddress{\state{MA}, \country{USA}}}
\address[4]{\orgdiv{Department of Psychiatry}, \orgname{University of Pittsburgh}, \orgaddress{\state{PA}, \country{USA}}}
\address[5]{\orgdiv{Department of Psychiatry and Behavioral Sciences}, \orgname{Emory University}, \orgaddress{\state{GA}, \country{USA}}}
\address[6]{\orgdiv{Department of Psychiatry and Behavioral Sciences}, \orgname{University of Washington}, \orgaddress{\state{WA}, \country{USA}}}
\address[7]{\orgname{Seattle Children's Research Institute}, \orgaddress{\state{WA}, \country{USA}}}
\address[8]{\orgdiv{Department of Psychiatry and Behavioral Neurosciences}, \orgname{Wayne State University}, \orgaddress{\state{MI}, \country{USA}}}

\corres{*Tugba Akkaya Hocagil, PhD, Department of Statistics and Actuarial Science, University of Waterloo, ON, N2V 2X5, Canada\\
\email{takkayahocagil@uwaterloo.ca\\
*Louise M. Ryan, PhD, School of Mathematical and Physical Sciences, University Of Technology Sydney, NSW, Australia\\
\email{Louise.M.Ryan@uts.edu.au}}}




\authormark{Akkaya Hocagil \textsc{et al}}

\abstract[Summary]{
Evidence from animal models and epidemiological studies has linked prenatal alcohol exposure (PAE) to a broad range of long-term cognitive and behavioral deficits. However, there is virtually no information in the scientific literature regarding the levels of PAE associated with an increased risk of clinically significant adverse effects. During the period from 1975-1993, several prospective longitudinal cohort studies were conducted in the U.S., in which maternal reports regarding alcohol use were obtained during pregnancy and the cognitive development of the offspring was assessed from early childhood through early adulthood. The sample sizes in these cohorts did not provide sufficient power to examine effects associated with different levels and patterns of PAE. To address this critical public health issue, we have developed a hierarchical meta-analysis to synthesize information regarding the effects of PAE on cognition, integrating data on multiple endpoints from six U.S. longitudinal cohort studies. Our approach involves estimating the dose-response coefficients for each endpoint and then pooling these correlated dose-response coefficients to obtain an estimated `global' effect of exposure on cognition. In the first stage, we use individual participant data to derive estimates of the effects of PAE by fitting regression models that adjust for potential confounding variables using propensity scores. The correlation matrix characterizing the dependence between the endpoint-specific dose-response coefficients estimated within each cohort is then run, while accommodating incomplete information on some endpoints. We also compare and discuss inferences based on the proposed approach to inferences based on a full multivariate analysis.}


\maketitle

\newpage
\input{sec1.tex}

\input{sec2.tex}

\input{sec3.tex}
\input{sec4.tex}

\input{sec5.tex}
\input{sec6.tex}
\input{sec7.tex}

\newpage
\input{Table1.tex}
\vspace{10mm}
\input{Table2.tex}
\vspace{10mm}
\input{Table3.tex}
\vspace{10mm}
\input{Table4.tex}
\vspace{10mm}
\input{Table5.tex}

\clearpage
\bibliographystyle{plain}
\bibliography{refs}
\clearpage
\input{appendix.tex}

\end{document}

%% file: sec1.tex
\section{Introduction}

Meta-analysis is a common approach used to combine quantitative evidence across studies to generate a global exposure/treatment effect that is more precise than estimates obtainable from individual studies alone.  Traditionally, this has been achieved by obtaining summary statistics from published studies and then combining those estimates using methods of meta-analyses. Although it is cost effective and easy to implement, meta-analyses tend to be underpowered and prone to ecological and confounding bias \citep{doi:10.1002/sim.2768, doi:10.1002/jrsm.4}. Individual patient data (IPD) meta-analysis provides a potential solution to avoid such biases, along with modeling flexibilities such as accounting for the correlation between endpoints and dealing with missing data at the patient level \citep{Riley2010MetaanalysisOI}. If one has access to the original patient level data, there may be a choice between a fully specified multivariate IPD  and a two-stage IPD approach. The full multivariate approach generally uses mixed-effects multilevel regressions to model between and within heterogeneity and quantify the effect of interest in a single model. Although this approach is considered flexible, it may be challenging for conducting and communicating the findings, especially regarding visualization with the hallmark forest plot. The alternative IPD approach involves modeling the data in two stages. In the first step, study-level estimates are obtained using separate regression models. In the second step, standard methods of meta-analysis are used to obtain an overall estimate. An important disadvantage of standard methods of meta-analysis is the assumption that effect size estimates being combined are independent. This assumption is violated when multiple outcomes are interest for synthesis across studies. To avoid the dependence of the effect sizes, several ad-hoc methods have been proposed including averaging the effect sizes and selecting one effect size per study. These ad-hoc approaches may lead to missed opportunities to utilize all available data to address the relevant research questions \citep{cheung-NR2019}. 

More principled approaches have been proposed to deal with correlated effects when conducting IPD meta-analysis.  These advances include multivariate meta-analysis which has been used to jointly synthesize the outcomes observed across studies to estimate multiple pooled effects simultaneously \citep{cite-key}. Another approach is  the three-level meta analytic model \citep{cheung2013, konstantopoulos-RSM2011,van-den-noortgate-BRM2013}, which has been used to account for dependence of effect sizes within clusters. This approach considers participants to contribute to only one effect size, so the non-independence is primarily introduced due to the nested structure of the effects \citep{cheung-NR2019}. 
An additional approach is based on a two-stage meta-analysis that uses summary measures. In this approach, dependency among effect sizes is handled via robust variance estimation (RVE) in which the dependence between the endpoints is not explicitly modeled, but instead the standard errors for the overall treatment effect or meta-regression coefficients are adjusted \citep{hedges-RSM2010}. This approach may require making a reasonable guess about the between outcome correlation to estimate the between-study variance and to approximate the optimal weights. 

In this paper, we propose an innovative approach: a hierarchical meta-analysis for the settings in which each cohort study provides multiple endpoints,  resulting in correlated sampling errors of the estimated effects. The work is motivated by a project that involves the integration of data from six longitudinal cohorts, each one using using multiple inter-related tests and assessment tools to measure child cognition. We assume that all endpoints are measuring an effect of exposure on a common underlying construct  A major strength of our approach is that it will enable us to pool data from diverse endpoints within each cohort and then assess the degree to which consistent patterns emerge across cohorts. 

In the proposed approach, we first derive the estimates and the standard errors by fitting univariate regression models for each separate outcome of interest. In the second stage, we  combine the summary measures  using a random-effects model. In this stage, unlike the existing methods of two-stage IPD analysis, we account for the correlated effect sizes within each cohort. First, some of the endpoints within a cohort were highly correlated causing unstable results. To overcome this problem, we obtain robust covariance estimates for effect size estimates within cohorts which we use at the second synthesis stage of the hierarchical meta-analytic approach which not all endpoints was observed for all children within a cohort. To address this issue, we derived a formula for the pairwise correlation between estimated effects using an adjustment, reflecting the fact that not all endpoints are observed for all children. To facilitate the creation of forest plots, downloading the results, and view information about the endpoints included in the analysis in an efficient way, we developed a web-based application which is written using the Shiny library and is hosted on a server using the open-source Shiny Server software (Rstudio and Inc., 2014).  In the last stage of our hierarchical meta analytic approach, we combine the independent effect size estimates obtained for each cohort in a random-effects model to obtain a global measure of the effect size across cohorts \citep{whitehead-book2002,10.1093/biomet/asq006}.

We compare and contrast the findings from proposed approach and a full multivariate analysis to answer a question that naturally arises when results of these two models coincide. Previous studies evaluated this question in different settings. Olkin and Simpsin \citep{olkin-sampson-BIOMETRICS1998} showed that in the case of comparing multiple treatments and a control with respect to a continuous outcome, the traditional meta-analysis based on estimated treatment contrasts is equivalent to the least squares regression analysis of individual patient data if there are no study-by-treatment interactions and the error variances are constant across trials. Mathew and Nordstorm \citep{mathew-nordstrom-BIOMETRICS1999} claimed that the equivalence holds even if the error variances are different across trials. Empirically, meta-analysis using original data has been found to be generally similar but not identical to meta-analysis using summary statistics. Whitehead and Lin and Zeng \citep{whitehead-book2002,10.1093/biomet/asq006}, showed that for all commonly used parametric and semi-parametric models, there is no asymptotic efficiency gain by analyzing original data if the parameter of main interest has a common value across studies, the nuisance parameters have distinct values among studies, and the summary statistics are based on maximum likelihood. More recently Kontopantelis \citep{doi:10.1002/jrsm.1303} conducted a comprehensive simulation study to compare one-stage and two-stage IPD analysis and concluded that a fully specified one-stage model is preferable especially when investigating interactions. We extend the results from these existing studies to the setting where there are correlated endpoints across multiple cohorts.

The remainder of the article is organized as follows. In Section 2, we introduce our motivating application based on a meta-analysis of correlated endpoints measures the effect of prenatal alcohol exposure (PAE) on cognition in six large cohort studies. In Section 3, we introduce notation and describe the two-stage analysis and modeling framework used to combine multiple correlated endpoints within a single cohort. In Section 4, we present the modeling framework used to combine pooled effect size estimates across cohorts. In section 5, we compare/contrast the proposed approach with the corresponding one-stage approach via simulation studies. In section 6, we illustrate our method using data from our motivating application.  Finally, in section 7, we discuss the strengths and limitations of our method.

%% file: sec2.tex
\section{Motivating Application}	
Evidence from animal models and epidemiological studies has linked PAE to a broad range of cognitive and behavioral deficits, growth impairment, and physical anomalies, which are known collectively as fetal alcohol syndrome disorders (FASD) \citep{doi:10.1097/01.ALC.0000145691.81233.FA,doi:10.1111/j.1530-0277.2007.00585.x,Cartere20160775,doi:10.1111/acer.14040}. Fetal alcohol syndrome (FAS), the most severe of the FASD, is characterized by distinctive craniofacial dysmorphology (small palpebral fissures, flat philtrum, thin vermillion), small head circumference, and growth retardation \citep{stratton-book1996}; \citep{hoyme-PEDIATRICS2005} while partial FAS (PFAS) is diagnosed in the presence of facial dysmorphology, a history of PAE and growth retardation, microcephaly, or central nervous system (CNS) impairment. Individuals with PAE who lack the characteristic pattern of dysmorphic features but exhibit cognitive and/or behavioral impairment are often given diagnosis of alcohol-related neurodevelopmental disorder (ARND), which is the most prevalent FASD.  Although the diagnosis of ARND requires a confirmed history of maternal alcohol consumption during pregnancy, there is virtually no information in the scientific literature regarding the levels of PAE associated with an increased risk of clinically significant adverse effects. In a recently updated set of clinical guidelines \citep{hoyme-PEDIATRICS2005}, risky drinking during pregnancy was defined as 6 or more drinks/week for a 2-week period or 3 or more drinks per occassion consumed on at least 2 occasions. However, in human studies, there is little empirical evidence of adverse effects at these levels of exposure.

Between 1975 and 1993, the National Institutes of Health (NIH) funded six longitudinal cohort studies in four U.S. cities (Seattle, Atlanta (two cohorts), Pittsburgh (two cohorts), and Detroit); these are described briefly in Section 1.2. In all but one of these studies, mothers were recruited and interviewed prospectively about their alcohol use during pregnancy and their children were followed longitudinally from infancy through young adulthood; the Atlanta 2 study \citep{doi:10.1002/dev.20215} recruited the mothers shortly following delivery, interviewed them about their drinking during pregnancy, and followed the children as in the other studies. All six studies had measured a variety of neurodevelopmental endpoints that assess five specific measures of cognitive function: IQ, executive function, academic achievement, and learning and memory. 

\subsection{The Six Cohort Studies}

\subsubsection{Seattle Cohort\normalfont{\citep{Streissguthetal1981}}  \hfill birth years: 1975-76}
All women who were enrolled in prenatal care by the 5th month of pregnancy at two large Seattle hospitals were eligible to participate. To  ascertain PAE,  participating mothers (N = 1529) were administered a Quantity-Frequency-Variability interview \citep{doi:10.15288/qjsa.1968.29.130} regarding alcohol, tobacco, and drug use for two time periods:  during pregnancy and just prior to pregnancy recognition. 462 newborns were selected based on an algorithm derived from maternal absolute alcohol (AA)/day, alcohol use/occasion, volume variability, and frequency of intoxication constructed to over-represent infants born to heavier drinkers. Controls included both abstainers and light drinkers. Infants were followed up at 8 and 18 months and 4,  7,  11,  14, 21,  25,  and 30 years. Although cohort retention was high (e.g., 82\% at 14 years), other children not initially selected whose mothers had been interviewed during pregnancy were added at  follow- 
up assessments to keep the sample size close to 500 at each examination.

\subsubsection{Atlanta Cohort 1\normalfont{\citep{article}} \hfill birth years: 1980-86} 
527 low socioeconomic status (SES), pregnant women were recruited at their first prenatal  visit at an urban Atlanta hospital serving a primarily African American, low income population. Women who reported drinking at least 1 oz AA/week during pregnancy were recruited. Nondrinkers, who were similar in demographic background, were recruited at the same time to serve as controls. Women were interviewed at recruitment about their alcohol and drug use; the majority reported drinking on weekends in a ‘binge‘ pattern. Infants were evaluated following birth. Sub-samples were followed up at 6 and 12 months and 7, 14, and 22 years.

\subsubsection{Atlanta Cohort 2\normalfont{\citep{Brownarticle}} \hfill birth years: 1992-1994}
306 mothers and their infants were recruited shortly after delivery at an urban Atlanta hospital.  111 reported having drunk alcohol during pregnancy, 71 of whom also had used cocaine (based on self-report or urine screen); 44 used cocaine but no alcohol; 151 did not drink alcohol or use cocaine. All participants were English speaking, 19 years or older, and had singleton births; most were African American and low SES. The infants were assessed at 2 and 8 years.

\subsubsection{Pittsburgh Cohort 1\normalfont{\citep{DAY1991329}}  \hfill birth years: 1983-86}
Participants were recruited from the prenatal clinic at a maternity hospital if they were English-speaking, age 18 or older,  and in their 4th  or 5th gestational month. A total of 1360 women were interviewed regarding their usual, maximum, and minimum consumption of beer, wine, liquor, tobacco, marijuana, and other illicit drugs prior to pregnancy and during the first trimester. Two cohorts were selected: (1) women who drank 3 or more drinks/week in the first trimester and a random sample of women who drank less or abstained, and (2) women who used marijuana at least twice/month in the first trimester and a random sample of those who used less or abstained.  Because women could be selected for either or both cohorts, there was a 48\% overlap. The cohorts were combined for this analysis: The birth sample consisted of 763 live singleton infants. The alcohol, tobacco, and drug use interview was repeated in the 7th gestational month and at delivery, when second and third trimester substance use information was obtained. The cohort consisted of women who were predominantly low income and of fairly equal numbers of Caucasian and African American women. Participants were followed up at 8 and 18 months, and 3, 6, 10, 14, 16, and 22 years.

\subsubsection{Pittsburgh Cohort 2\normalfont{\citep{Galearticle}} \hfill birth years: 1988-93}
English-speaking women in their 4th or 5th month of pregnancy attending the prenatal clinic at a large inner-city hospital who were 18 years old or older were interviewed regarding their usual, maximum, and minimum consumption of cocaine, alcohol, marijuana, tobacco, and other drugs prior to pregnancy and during the 1st trimester. Every woman who reported any cocaine/crack use during the first trimester was enrolled in the study cohort, as was the next woman interviewed who reported  no cocaine or crack use during both the year prior to pregnancy and the first trimester. Although crack/cocaine use was the criterion for recruitment, a large proportion of these women also drank moderate-to-heavy levels of alcohol; some of the adverse effects subsequently seen on development were independently attributable to the alcohol exposure. The alcohol and drug use interview was repeated at the  end of the second and third trimesters, and offspring were assessed at 1, 3, 7, 10, 15, and 21 years.   The birth cohort consisted of 295 women and infants; the women were predominantly of low socioeconomic status and were roughly equally divided by Caucasian and African American race.  

\subsubsection{Detroit Cohort\normalfont{\citep{doi:10.1111/j.1530-0277.1993.tb00744.x}} \hfill birth years: 1986-89}
All women (N > 6400) enrolling in the antenatal maternity clinic at a large, inner-city hospital were interviewed regarding their alcohol use at their first prenatal visit (M = 23.4 wk gestation; SD=7.9), using a timeline follow-back interview \citep{jacobson-PREDIATRICS2002}. The mother provided detailed information about her alcohol consumption on a day-by-day basis during the previous 2 weeks with recall linked to specific times of day and activities and was also asked to recall her day-by-day drinking during a typical week around time of conception. Moderate and heavy drinking women were overrepresented in the sample by including all women reporting at least 0.5 oz AA at conception and a random sample of approximately 5\% of the lower level drinkers and abstainers. The 2-week timeline follow-back interview was repeated at each prenatal clinic visit (M = 5.4 visits). To reduce the risk that alcohol might be confounded with cocaine exposure, 78 heavy cocaine (<2 days/wk), light alcohol (<7 drinks/wk) users were also included in the final sample, which consisted of 480 pregnant women and their children. Participants were followed up at 6.5, 12, and 13 months and 7, 14, and 19 years.

%% file: sec3.tex
\section{Notation and Model Formulation}

Let $Y_{ijk}$ be the random variable representing response $k$ for individual $j$ in cohort $i$,
$k = 1, \ldots, K$, $j = 1, \ldots, J_i$, where $J_i$ is the number of individuals in cohort $i$, $i = 1, \ldots, I$. 
Let $A_{ij}$ be the exposure of interest (i.e. prenatal alcohol exposure) for individual $j$ in study $i$
and $S_{ij}$ be their corresponding propensity score. 

\begin{equation} \label{eq-Y-linear-model-separate}
Y_{ijk} = \alpha_{ik} + B_{ik} \, A_{ij} + \gamma_{ik} \, S_{ij} + E_{ijk} \,,
\end{equation}
\noindent
where $B_{ik}$ is the effect of a one-unit increase in $A_{ij}$ (alcohol volume) on the mean for response $k$ in cohort $i$
given the propensity score $S_{ij}$, $j = 1, \ldots, J$, $k = 1, \ldots, K$, $i=1, \dots, I$.
The propensity score is calculated separately for each cohort in our setting 
since the sets of covariates measured differ between cohorts \citep{Hwang2020}. 
We assume here that conditioning on the propensity score renders the exposure variable independent of all confounders and so that it is sufficient to condition on $S_{ij}$ rather than the confounders themselves \citep{doi:10.1198/016214504000001187}. 

The parameter $\gamma_{ik}$ characterizes the effect of the propensity score for a given level of alcohol exposure
and $E_{ijk}$ is the error term which has mean zero and variance $\sigma^2_{ik}$, $k = 1, \ldots, K$, $i=1, \dots,I$.

We suppose that the effects $B_{ik}$ vary about some average exposure effect, with, 
\begin{equation} \label{eq-average-exposure-Bk}
B_{ik} \sim N(\beta_{i}, \phi_{i}) \,,
\end{equation}
\noindent
where $\beta_i$ is the exposure effect for cohort $i$ and $\phi_i$ represents the heterogeneity of the response-specific exposure effects for cohort $i$. 

Across cohorts, we suppose that the average exposure effects $\beta_{i}$ vary about some average exposure effect, with, 
\begin{equation} \label{eq-average-exposure-Bk_2}
\beta_i \sim N(\beta_{o}, \eta^2) \,,
\end{equation}
\noindent
where  $\beta_{o}$ represents the ``average effect'' of a one-unit increase in the exposure across all cohorts, and is our parameter of ultimate interest.
The variance $\eta^2$ in (\ref{eq-average-exposure-Bk_2}) reflects the extent of heterogeneity of the cohort-specific exposure effects. 

In the next two subsection, we describe a two-stage approach to estimation and inference with data from a single cohort and in Section 4, we show how to synthesize cohort-specific exposure effects to obtain an estimate for the average effect of a one-unit increase in the exposure across all cohorts. 

\subsection{Stage I Estimation for a Single Cohort}
In this section, we temporarily omit the subscript $i$ and describe a two-stage approach to estimate the average exposure effect for a single cohort where the effects are correlated. 
Before model fitting we standardize the responses so that they have the same first two moments as the Full-Scale IQ variable 
which has a mean of 100 and a standard deviation of 15.
By conducting this standardization the exposure effects can be expressed  
in terms of the decrement in IQ  associated with a one-unit increase 
in prenatal alcohol exposure \citep{axelrad-EHP2007}. 

For the first stage, we fit separate linear models for each response, assuming 
\begin{equation} 
Y_{jk} = \alpha_{k} + B_{k} \, A_{j} + \gamma_{k} \, S_{j} + E_{jk} \,,
\end{equation}
\noindent
where $B_{k}$ is the effect of a one-unit increase in $A_{j}$ (alcohol volume) on the mean for response $k$,
given the propensity score $S_{j}$, $j = 1, \ldots, J$, $k = 1, \ldots, K$.
The parameter $\gamma_{k}$ characterizes the effect of the propensity score for a given level of alcohol exposure
and $E_{jk}$ is the error term which has mean zero and variance $\sigma^2_{k}$, $k = 1, \ldots, K$.

We suppose that the effects $B_{k}$, $k = 1, \ldots, K$
vary about some average exposure effect, with
\begin{equation} 
B_{k} \sim N(\beta, \phi),
\end{equation}
\noindent
independently and identically distributed and $\beta$ is the average exposure effect. The variance $\phi$ reflects the extent of heterogeneity of the response-specific exposure effects for a single cohort. 


If we let $X_{jk} = (1, A_{j}, S_{j})'$ be the covariate vector, we can be write
\begin{equation} \label{eq-Y-giv-BE-linear-model-compact}
Y_{jk} = X'_{jk} \, \theta_{k} + E_{jk} 
\end{equation}
where $\theta_{k} = (\alpha_{k}, B_{k}, \gamma_{k})'$.
We assume $E_{jk}\independent (A_{j}, B_{k},S_{j})$ with $E_{jk} \sim N(0, \sigma_{k}^2)$ are i.i.d. for $j = 1, \ldots, J$, $k=1,2,\ldots, K$.
Note that $X_{jk}$ does not vary by response type 
since the exposure variable $A_{j}$ and the propensity score $S_{j}$ are individual level covariates,
but we retain this notation for generality. 

We next define $K \times 1$ vectors $Y_j = (Y_{j1}, Y_{j2}, \ldots, Y_{jK})'$,
$\alpha = (\alpha_1, \ldots, \alpha_K)'$,
$B = (B_1, \ldots, B_K)'$ and $\gamma = (\gamma_1, \ldots, \gamma_K)'$ and a $K \times 3K$ covariate matrix 
\begin{equation}
X'_j = \left[
\begin{array}{ccccc}
X'_{j1} & 0          & 0 & \cdots & 0 \\
0          & X'_{j2} & 0 & \cdots & 0 \\
\vdots   &            & \ddots &   & 0 \\
0          & \cdots  & 0         & 0 & X'_{jK} \\
\end{array}
\right] \,,
\end{equation}
\noindent
where the $0$s in this matrix refer to $3\times 1$ vectors of $0$s. The model given by (\ref{eq-Y-giv-BE-linear-model-compact}) can then be represented in a unifying model
\begin{equation}
Y_j = X'_j \, \theta + E_j 
\end{equation}
\noindent
where 
$\theta = (\theta'_1, \ldots, \theta'_K)'$ is a $3 K \times 1$ vector of parameters,
$E_j = (E_{j1}, \ldots, E_{jK})'$ and $E_j \sim N(0, \Sigma)$, 
where $\Sigma$ is a $K \times K$ covariance matrix with diagonal entries 
$\Sigma_{kk} = \sigma_k^2 = {\rm var}(E_{jk})$.  
The off-diagonal entries 
$\Sigma_{kl} = \sigma_{kl}={\rm cov}(E_{jk}, E_{j \ell})$ 
accommodate a conditional dependence (given $X_{jk}$, $X_{jl}$, $B$) between the responses from the same individual. 


In the second stage of estimation these estimates are pooled over 
to obtain a single estimate of the global measure of the causal effect denoted by $\beta$ in (\ref{eq-average-exposure-Bk}). We begin the second stage by estimating the covariance between the errors  ${\rm cov}(E_{jk}, E_{jl}) = \sigma_{kl}$, $l \ne k = 1, \ldots, K$, characterizing the dependence between the stage I estimators $\widehat{\theta}_1, \ldots, \widehat{\theta}_K$

The challenge in estimating the covariance between the errors  ${\rm cov}(E_{jk}, E_{jl}) = \sigma_{kl}$, $l \ne k = 1, \ldots, K$ is
that not all individuals contribute data for all responses. To accommodate this fact, we introduce the indicators $R_{jk} = I(Y_{jk}~\mbox{is observed})$, $k = 1, \ldots, K$, and assume that the responses are missing at random (MAR) according to Little and Rubin (2019).  

If $S_{jk}(\theta_{k})=X_{jk}(Y_{jk}-X_{jk}' \, \theta_{k})$ 
is the desired contribution from individual $j$ to the score function for $\theta_{k}$
given $B$,
the observed data score equation for estimating $\theta_{k}$ at stage I can be written as
\begin{equation} \label{eq-stageI-Sk-thetak}
S_{k}(\theta_{k})= \sum_{j=1}^{J} R_{jk} S_{jk}(\theta_{k}) = 0 \,,
\end{equation}
the solutions to which are
\begin{equation}
\widehat{\theta}_{k}=\sum_{j=1}^{J} R_{jk} \, (X_{jk} \, X'_{jk})^{-1} \, X_{jk} \, Y_{jk} \,, \quad k = 1, \ldots, K \,.
\end{equation}
\noindent
The maximum likelihood estimate of $\sigma^2_{k}$ is
\begin{equation} \label{eq-mle-sigma2k}
\widehat{\rm var}(E_{jk})
= \widehat{\sigma}^{2}_{k}
=\dfrac{\sum_{j=1}^{J} R_{jk} \, (Y_{jk}-X'_{jk} \, \widehat{\theta}_{k})^2}{n_k}
\end{equation}
where $n_k=\sum_{j=1}^{J} R_{jk}$ 
is the number of individuals contributing to the estimation of $\theta_{k}$, $k=1,2,\dots, K$.
We also obtain the maximum likelihood covariance estimate as
\begin{equation} \label{eq-mle-sigmakl}
\widehat{\rm cov}(E_{jk}, E_{jl})
=\widehat{\sigma}_{kl}
= \dfrac{\sum_{j=1}^{J} R_{jk} \, R_{jl}  \,(Y_{jk}- X'_{jk} \, \widehat{\theta}_{k}) \,  (Y_{jl}- X'_{jl} \, \widehat{\theta}_{l})}
            {n_{kl}}
\end{equation}
\noindent
where $n_{kl} = \sum_{j=1}^J R_{jk} \, R_{jl}$,
which is consistent
under a missing at random assumption \citep{little-rubin-book2019}.  
We then let $\widehat{\Sigma}$ denote the estimated covariance matrix for the errors 
where $\sigma_{kl}$ are adjusted for missing data as

\begin{equation} 
\widehat{\rm cov}(E_{jk},E_{jl})_{adjusted}
=\dfrac{\widehat{\sigma}_{kl} \,n_{kl}}
{n_{k}\,n_{l}}
\end{equation}
\noindent

where $n_k=\sum_{j=1}^{J} R_{jk}$ and  $n_l=\sum_{j=1}^{J} R_{jl}$
are the number of individuals contributing to the estimation of $\theta_{k}$ and $\theta_{l}$ respectively. 

More details regarding the derivation of the dependence between the estimates $\widehat{\theta}_{k}$ and $\widehat{\theta}_{l}$ 
are provided in the appendix.

\subsection{Stage II: Synthesis across Responses within a Cohort}
\label{SEC3.2}

To consider the synthesis of estimators across all responses
we note that
\begin{equation*}
E(\widehat{B}_k - \beta) = E \{ (\widehat{B}_k - B_k) + (B_k - \beta) \} = 0
\end{equation*}
\noindent
so $\widehat{B}$ is comprised of $K$ dependent unbiased estimators of $\beta$.
Thus
\begin{equation} \label{eq-Bhat-MVN}
\widehat{B} \sim {\rm MVN}(\mu(\beta), \Psi(\phi))
\end{equation}
\noindent
asymptotically,
where $\mu(\beta)$ is a $K \times 1$ vector with each element equal to $\beta$.
If we let $\widehat{\Psi}(\phi) = J^{-1} \, \widehat{\Gamma} + \Delta \, \phi$
then based on (\ref{eq-Bhat-MVN}) we may specify a pseudo-likelihood $PL(\beta, \phi)$ for $(\beta, \phi)$ given by
\begin{equation}\label{eq-PL-beta-tau2}
PL(\beta, \phi) 
\propto \dfrac{1}{ (2 \pi)^{K/2} \, \sqrt{\widehat{\Psi}(\phi)|}} 
\exp \left(-\frac{1}{2} \, (\widehat{B}- \mu(\beta))'  \, \widehat{\Psi}^{-1}(\phi)  \, (\widehat{B} - \mu(\beta))  \right) \,.
\end{equation}

Note that (\ref{eq-PL-beta-tau2}) could be maximized with respect to $(\beta, \phi)$,
but we proceed in a computationally convenient iterative approach based on (\ref{eq-PL-beta-tau2}).
Given an estimate $\phi^{(r)}$ 
we compute an estimate $\beta^{(r)}$ based on a linear combination of 
$\widehat{B}_1, \ldots, \widehat{B}_K$.
The most efficient linear estimator of $\beta$ has the form
\begin{equation} \label{eq-linear-estimator-beta-hat}
\widehat{\beta} = [\mathbbm{1}' \, [\Psi(\phi)]^{-1} \, \widehat{B}] / [ \mathbbm{1}' \, [\Psi(\phi)]^{-1}  \, \mathbbm{1} ] \,,
\end{equation}
\noindent

so we replace $\Psi(\phi)$ with an estimator 
$\widehat{\Psi}(\phi^{(r)}) = J^{-1} \, \widehat{\Gamma} + \Delta \, \phi^{(r)}$.
We could invert $\widehat{\Psi}(\phi^{(r)})$ but in practice it may be difficult and while a generalized inverse could be used,
the weights resulting from this approach were often found to vary greatly in magnitude and even in sign. 
We therefore adopted an alternative more stable linear estimate using inverse variance weights,
whereby we replace $\Psi(\phi)$ with
$diag(J^{-1}  \, \widehat{\Gamma}_{kk} + \phi^{(r)} \,,~k = 1, \ldots, K)$ in (\ref{eq-linear-estimator-beta-hat}) to obtain $\widehat{\beta}^{(r)}$.
We then maximize $PL(\widehat{\beta}^r, \phi)$ with respect to $\phi$ to obtain $\phi^{(r+1)}$, with which we recompute $\widehat{\beta}^{(r+1)}$ 
and repeat iteratively until convergence;
we let $(\widehat{\beta}, \widehat{\phi})$ denote the estimates upon convergence.
A robust variance estimate is then obtained for $\widehat{\beta}$ based on 
$\widehat{\Psi}(\widehat{\phi})$ which is given by
$\widehat{\rm var}(\widehat{\beta}) = [ \mathbb{I}' \, \widehat{\Psi}(\widehat{\phi}) \, \mathbb{I}]^{-1}$.
When we consider data from multiple cohorts we reintroduce the subscript $i$ and write the corresponding estimates for cohort $i$ as $\widehat{\beta}_i$ and
\begin{equation}
\widehat{V}_i(\widehat{\beta}_i) = \widehat{\rm var}(\widehat{\beta}_i) = [ \mathbb{I}' \, \widehat{\Psi}_i(\widehat{\phi_i}) \, \mathbb{I}]^{-1}
\end{equation}
respectively, $i=1,\ldots, I$.

\subsection{An Alternative One-Stage (Fully Specified Multivariate) Approach}
Note that the parameters estimated according to the hierarchical meta-analytic approach can alternatively be fitted in one-step via software for fitting hierarchical mixed effect linear models.
To do so we define $K-1$ covariates $T_{jkr} = I(k = r)$, $r = 2, \ldots, K$ which indicates the outcomes is for response $k$, $k = 2, \ldots, K$.
We may put these $K-1$ indicators in vector format and define the $(K-1)\times 1$ vector $T_{jk} = (T_{jk2}, \ldots, T_{jkK})'$.
We consider the response for verbal IQ as the reference type and let $T_{ik2} = 1$ for performance IQ,
$T_{ik3} = 1$ for freedom from distractibility (Table 2).
Then we fit the model
\begin{equation*}
Y_{jk}  = \alpha_1 + B_k \, A_j + \gamma_1 \, S_j + \tau' \, T_{jk} + \zeta' \, S_j \, T_{jk} + E_{jk} \,, \quad k = 1, \ldots, K \,.
\end{equation*}
where $\tau=(\tau_2, \ldots, \tau_K)'$ and 
$\zeta = (\zeta_2,\ldots, \zeta_K)'$ are $(K-1) \times 1$ vectors with 
$\tau_r = \alpha_r - \alpha_1$ and 
$\zeta_r = \gamma_r - \gamma_1$, $r=2,\ldots, K$.
We assume $B_k \sim N(\beta, \phi)$ as specified in (\ref{eq-average-exposure-Bk}) where $B_k$ is the effect of a one-unit increase in $A_j$ on the 
mean of response $k$ given the propensity score $S_j$,
$\beta$ is the parameter of ultimate interest representing the ``average causal effect'' of a one-unit increase in the exposure across 
all responses within the cohort, and 
$\phi$ characterizes the degree of heterogeneity in the effect across responses. 
We also assume $E_j = (E_{j1}, \ldots, E_{jK})'$ is a $K \times 1$ error term with 
$E_j \sim {\rm MVN}(0, \Sigma)$
with $\Sigma$ a $K \times K$ covariance matrix as in Section 3.1.
The one-step approach involves simultaneous estimation of all fixed effects, $\beta$, $\phi$ and $\Sigma$ at once.
This can be fitted using software for fitting hierarchical linear mixed effects models.
We let 
$\tilde{\beta}_i$ denote the estimate of $\beta_i$ obtained from fitting the hierarchical model to the data from cohort $i$ and
$\tilde{V}_i(\tilde{\beta}_i)$ denote the correspondign variance estimate based on the observed information matrix, $i=1,\ldots, I$.

%% file: sec4.tex
\section{Synthesis across Cohorts}
In the previous section, we described methods for synthesizing data across multiple outcomes to obtain estimates of the global causal effect based on a 
two-stage approach and based on fitting a hierchical mixed effect model.
These methods were based on analyzing data from a single cohort. Here, we describe how to combine cohort-specific estimates to obtain a an overall estimate
of a causal effect while accommodating possible heterogeneity.
The approach described in Section 3.2 is an extension of the approach described by Viechtbauer and implemented in the metafor package \citep{metaforR} which deals with 
independent estimates; Section 3.2 adapted the methods to deal with dependent effect estimates so what follows is a simplification of the approach for the 
last stage of the data synthesis.  
We describe it briefly as follows. 

We consider $\beta_i$ as the global causal effect of exposure in cohort $i$ reflecting the impact of an increment in the volume of  prenatal alcohol exposure 
on the common underlying construct; we let $\hat\beta_i$ be the corresponding estimate.
Note that the studies draw individuals from different populations and so the composition of the samples varies across cohorts.
Moreover, the methods used to measure exposure along with the precise nature of the outcomes differ between studies, even though they were measuring the same latent attributes regarding cognition.
We therefore wish to accommodate a component of variation between studies (heterogeneity) for the true effects which we accomplished by use of a random effects model of the form
\begin{eqnarray}
\hat{\beta}_{i} & = & \beta_{i}+\epsilon_{i}\\
\beta_{i} & = & \beta_\circ + u_{i}
\end{eqnarray}
where 
we let $\epsilon_i \sim N(0, \widehat{V}_i(\widehat{\beta}_i))$ reflect the sampling variation of the estimator from cohort $i$ about the true effect $\beta_i$, 
and $u_{i}\sim N(0,\eta^2)$ reflects the heterogeneity of the global cohort-specific causal effects across studies. 
The parameter $\beta_\circ$ represents the overall global effect which is the parameter of ultimate interest.
Through this variance decomposition then upon introducing the heterogeneity between studies we have 
$\widehat{\rm var}(\widehat{\beta}_i)=  
\widehat{V}_i(\widehat{\beta}_i) + \eta^2$.
The synthesis is achieved in a simlar spirit to Section 3.2 whereby we consider a pseudo-likelihood of the form
\begin{equation}
\label{eq-PL-beta-eta2}
PL(\beta_\circ, \eta)
\propto 
\prod_{i=1}^I \left\{ \dfrac{1}{ (2 \pi)^{I/2} \, \sqrt{(\widehat{V}_i(\widehat{\beta}_i) + \eta^2)}}
\exp \left( - \frac{(\widehat{\beta}_i- \beta_\circ)^2}{2 ( \widehat{V}_i(\widehat{\beta}_i) + \eta^2)} \right) \, \right\}.
\end{equation}
The pooled exposure effect estimate $\widehat{\beta}_\circ$ is obtained as a weighted average of the $\widehat\beta_i$ terms 
with cohort weights equal to the inverse of $\widehat{V}(\widehat{\beta}_i)+\hat\eta^2$ where $\hat{\eta}^2$ is obtained as the solution to iteratively 
maximizing (\ref{eq-PL-beta-eta2}).
The R package `metafor' can be used to carry out this final stage of the data synthesis.
If the linear model of Section 3.3 is used for simultaneous estimation of the overall causal effect then $\tilde{\beta}_i$ and 
$\tilde{V}_i(\tilde{\beta}_i)$  can be used in a similar fashion to obtain the estimator $\tilde{\beta}_o$.

%% file: sec5.tex
\section{Simulation Studies}
For the simulation studies, we consider $k$ correlated continuous endpoints from a single study. We generated outcomes from the following linear regression model: 

\begin{equation} 
Y_{jk} = \alpha_k + \beta_k \,X_j + \gamma_k \, Z_j + E_{jk} \,,
\end{equation}

where $Y_{jk}$ be the random variable representing response k for individual j, k = 1,...,K, j = 1,...,J, and  $\beta_k$ is the effect of a one-unit increase in $X_j$ on the mean for response $k$ given the covariate $Z_j$. We let $\beta_k$ vary about some average exposure effect within a study, with $\beta_k \sim N(\beta, \tau^2)$. The parameter $\gamma_k$ characterizes the effect of the covariate for a given level of exposure, $X_{j}$. We also assume $E_{j} = (E_{j1}, \dots, E_{jK} )$ is a $K \times 1$ error term with $E_{j}\sim MVN(0, \Sigma)$ with $\Sigma$ a $K \times K$ covariance matrix. The simulations were performed in different scenarios. We generated the effect size for the exposure, $\beta_k$ from normal distribution with mean 3 and variance $\tau^2$.  Scenarios were created by manipulating the number of outcomes (k) and, varying between-study heterogeneity ($\tau^2$). We consider the scenarios where the number of outcomes is equal to 3, 5, and 10 and $\tau^2$ takes the values 0.10,0.25 and 0.50.
For each combination of the simulation parameters, we generated 1000 datasets with the sample size of 500 for each endpoint. For each dataset, we performed the two types of meta-analysis, i.e. the one based on the proposed approach versus the full multivariate analysis.  

We evaluated the performance of the proposed approach in simulation settings previously described, over 1000 iterations. The estimates of interest were the average exposure effect $\beta$. To allow for a comprehensive comparison, performance was assessed on a range of metrics: empirical mean bias (EBIAS), average model based standard error (ASE), empirical standard error (ESE) and coverage probability (CP).  The results are summarized in Table 1. Patterns of empirical mean bias were very low and comparable for the two methods, with the exception of larger between heterogeneity and smaller number of outcomes. In those particular scenarios, the proposed approach was the best performer. Coverage probability for the proposed method was about to the nominal 0.95 for all scenarios considered in this paper . 

%% file: sec6.tex
\section{Prenatal Alcohol Exposure and Cognitive Function in Children}
For this paper we used data from six large cohort studies to assess the effects of PAE on IQ, which is a measure of cognitive function. The proposed hierarchical meta-analytic approach is well-suited to assess the effect of PAE on IQ measure since it enables us to pool data from diverse, correlated endpoints across cohorts. Table 2 lists the endpoints by cohort considered in this paper. To yield sufficiently precise estimates of effect size, we considered a broad set of potential confounders when fitting separate linear models for each endpoint. Since each cohort provided a somewhat different set of control variables, we employed a propensity score approach to adjust for potential confounders (\citealp{AkkayaHocagil_2020}). We estimated the propensity score for each cohort separately and included the propensity score in the linear model as an additional covariate as in model (3.1). 

For each outcome k, $\beta_{k}$ was estimated from model (3.1). Table 3 lists the estimated effect size and standard errors from the first stage of the hierarchical meta-analytic approach. The aim of the second stage of the proposed methods is to pool the estimates of the PAE and estimate the cohort specific overall true mean effect $\beta_{i}$ while adjusting for the fact that endpoints are correlated within a cohort and accommodating incomplete information on some endpoints. Table 4 shows the estimated effect sizes and standard errors for each cohort. To compare and contrast the results obtained from our method, we conducted a fully specified multivariate analysis to estimate a pooled effect size for each cohort study using SAS procedure `proc mixed` . Table 4 shows the estimated effect sizes for each cohort obtained from the fully specified multivariate model. Both methods provided impressively similar estimates for the effect sizes and the standard errors. Although the difference was not substantial, these two methods provided slightly different estimates for the between endpoints heterogeneity.

To combine the independent effect size estimates across cohorts and obtain a global effect size estimate of PAE on IQ at age 7 years, we used the R package ``metafor" to pool the estimates resulting from our hierarchical meta-analysis and the fully specified multivariate model. Table 5 provides the estimated effect size of the PAE on childhood IQ across cohorts. The resulting global effect sizes from the two methods were almost identical. 

%% file: sec7.tex
\section{Discussion}

	In this paper, we have proposed an extension of the standard procedure for two-stage IPD meta-analysis. Our procedure follows the same steps as conventional methods for two-stage IPD analysis for making inference about the effect size but extends these analyses by accounting for the correlation between endpoints and by accommodating incomplete data on some endpoints. 
	
	Our approach has several advantages over the one-stage IPD meta-analysis. Firstly, it builds upon the two-stage IPD meta-analysis that practitioners are already familiar with. Secondly with our approach, one can create forest plots to visualize the estimated effect sizes for each endpoint. Thirdly, our approach is less likely to encounter convergence problems compared to the one-stage IPD meta-analysis. Finally, our approach uses the known within study variances, which helps obtain more precise estimates.
	
	We evaluated and compared our approach with a fully specified multivariate analysis. In simulation scenarios considered in this paper, we observed that the proposed approach can successfully reduce bias relative to the fully specified multivariate approach. Our simulation results suggest that, when the number of endpoints is small and the between endpoints variance is large, our proposed approach outperforms the multivariate analysis.
	
    We illustrated our approach using data on childhood IQ from six cohorts. We analyzed 18 outcomes from the six cohorts. For our data application, it was important to adjust for the fact that endpoints within a cohort are correlated and some of the endpoints have incomplete data. To obtain a global effect size estimate, we conducted a hierarchical meta-analysis. In the first stage, we obtained effect size estimates for each endpoint separately. In the second stage, we employed our proposed approach to obtain cohort-specific pooled effect size estimates while adjusting for between-endpoint correlation and incomplete data. In the last stage, we combined effect sizes across the cohorts employing a random-effects model. We compared the results from our approach with the results from the fully specified multivariate approach. In this comparison, our method performed well and thus provides an useful innovative tool for performing and interpreting meta-analyses with the correlated effect sizes.

%% file: Table1.tex
\begin{sidewaystable}
\centering
\caption{Results of a simulation study assessing the performance of our hierarchical meta-analysis and a full multivariate analysis in a variety of settings}
  \begin{tabular}{lccccccccc}
 &  &  \multicolumn{4}{c}{Hierarchical Meta-Analytic Approach} &  \multicolumn{4}{c}{One-stage (full multivariate) approach} \\
\hline \\
 $\tau^2$ & k & EBIAS & ASE & ESE & CP & EBIAS & ASE & ESE & CP \\
0.10 & & &  & & & &  &  & \\ 
& 10 & 0.0070 & 0.16 & 0.14 & 0.96 & 0.0000 & 0.10 & 0.14 & 0.92\\
 & 5 & 0.0060 & 0.20 & 0.20 & 0.96 & 0.0001 & 0.13 & 0.14 & 0.99 \\
 & 3 & 0.0090 & 0.23 & 0.25 & 0.95 &  0.0009& 0.14 & 0.14 & 0.93  \\
 0.25 &  & &  &  & & & & & \\
 &10 & 0.0003 & 0.11 & 0.10 & 0.95 & 0.02 &   0.18 & 0.24 & 0.88 \\
 & 5 & 0.0002 & 0.13 & 0.13 & 0.95 & 0.03 & 0.19 & 0.21 & 0.91 \\ 
 & 3 & 0.0100 & 0.15 & 0.15 & 0.95 & 0.02 & 0.21 & 0.23 & 0.94\\
 0.50 &  &  &  &  &  &  &  & & \\
 & 10 & 0.0040 & 0.16 & 0.15 & 0.95 & 0.04 & 0.26 & 0.29 & 0.96 \\
 & 5 & 0.0090 & 0.21 & 0.20 & 0.94 & 0.06 & 0.27 & 0.30 & 0.97 \\
 & 3 & 0.0100 & 0.23 & 0.24 & 0.94 &  0.12 & 0.31 & 0.34 & 0.92 \\

 \\
 \hline
\end{tabular}
\begin{tablenotes}
\centering 
\item EBIAS: Empirical bias, ASE: Average model based standard error, ESE: Empirical standard error, CP: Coverage probability
\end{tablenotes}
\end{sidewaystable}

%% file: Table2.tex
\begin{table} [bt]
\centering 
\caption{IQ Related Outcomes Assessed at Age 7 in the Six Cohorts}
\begin{threeparttable}
\begin{tabular}{lll}
{Cohort} & {Endpoints}\\
  {Seattle}&\\
 &WISC Verbal IQ\\
 &WISC Performance IQ\\
 {Atlanta 1}&\\
 &Kaufman ABC Simultaneous processing\\
 &Kaufman ABC Sequential processing\\
 {Atlanta 2}&\\
 &DAS Verbal standard score\\
 &DAS Nonverbal standard score\\
 &DAS Spatial standard score\\
 {Pittsburgh 1}&\\
 &Stanford-Binet Verbal reasoning\\
 &Stanford-Binet Abstract reasoning\\
 &Stanford-Binet  Quantitative reasoning\\
 &Stanford-Binet  Short-term memory\\
 {Pittsburgh 2}&\\
 &Stanford-Binet Verbal reasoning\\
 &Stanford-Binet Abstract reasoning\\
 &Stanford-Binet Quantitative reasoning\\
 &Stanford-Binet  Short-term memory\\
 {Detroit}&\\
 &WISC Verbal IQ\\
 &WISC Performance IQ\\
 &WISC Freedom from distractibility\\
\hline  
\end{tabular}
\end{threeparttable}
\end{table}





%% file: Table3.tex
\begin{table} [htbp]
\centering 
\caption{Summary statistics of IQ related outcomes assessed at age 7}
  \begin{tabular}{llll}
     \hline
     \hline
  \multicolumn{4}{c}{Stage I}\\
     \hline
Cohort & Response type & Estimated effect size & SE \\
Detroit &	WISC Verbal IQ & -4.2 & 3.2\\
Detroit &	WISC Performance IQ &	-3.7 & 3.2\\
Detroit & WISC Freedom from distractibility& -10.3 & 3.1\\
Seattle & WISC Verbal IQ & -0.5 & 2.6\\
Seattle & WISC Performance IQ & -1.9 & 2.6\\
Atlanta  Cohort 1 & Kaufman ABC Simultaneous processing & -6.9 & 2.9\\
Atlanta  Cohort 1 & Kaufman ABC Sequential processing	 & -1.9 & 2.9\\
Atlanta Cohort 2 & DAS Verbal standard score & -5.9 & 3.2\\
Atlanta Cohort 2 & DAS Nonverbal standard score & 1.7 & 3.3\\
Atlanta Cohort 2 &  DAS Spatial standard score &-0.9 & 3.3\\
Pittsburgh Cohort 1 & Stanford Binet Verbal reasoning & -5.8 & 3.0\\
Pittsburgh Cohort 1 & Stanford Binet Abstract reasoning & -5.0 & 3.0\\
Pittsburgh Cohort 1 & Stanford Binet Quantitative reasoning & -1.9 & 3.0\\
Pittsburgh Cohort 1 & Stanford Binet Short term memory	& -5.3 & 3.0\\
Pittsburgh Cohort 2 & Stanford Binet Verbal reasoning & -0.3 & 3.1\\
Pittsburgh Cohort 2 & Stanford Binet Abstract reasoning & -1.8	& 3.0\\
Pittsburgh Cohort 2 & Stanford Binet Quantitative reasoning & -1.1& 3.1\\
Pittsburgh Cohort 2 & Stanford Binet Short term memory	& -3.5 &3.1\\
       \hline
\hline
\end{tabular}
\end{table}

%% file: Table4.tex
\begin{table} [htbp]
\centering 
\caption{Pooled effect size estimates of prenatal alcohol exposure for each cohort}
  \begin{tabular}{lllllll}
     \hline
     \hline
     &\multicolumn{6}{c}{Stage II}\\
&\multicolumn{3}{c}{Hierarchical Meta Analytic Approach} & \multicolumn{3}{c}{Full Multivariate Approach}\\
     \hline
Cohort & Effect Size  & SE & $\widehat\tau^2$ & Effect Size  & SE & $\widehat\tau^2$ \\
Detroit & -6.1 & 3.2 & 7.8 & -6.1 & 3.1 & 6.2 \\
Seattle & -1.2 & 2.3 & 0.0 & -1.2 & 2.3 & 0.0 \\
Atlanta Cohort 1 & -4.4 & 3.0 & 5.1 & -4.4 & 3.4 & 6.2 \\
Atlanta Cohort 2 & -1.9 & 3.0 & 7.8 & -2.0 & 3.2 & 8.2  \\
Pittsburgh Cohort 1 & -4.3 & 2.4 & 0.0 & -4.0 & 2.6 & 0.0\\
Pittsburgh Cohort 2 & -1.6 & 2.5 & 0.0 & -1.6 & 2.5 & 0.0 \\
       \hline
\hline
\end{tabular}
\end{table}

%% file: Table5.tex
\begin{table}
\centering
\caption{Estimated effects size of prenatal alcohol exposure on IQ at age 7}
  \begin{tabular}{llll}
  & Stage III\\
Method &Global effect size &  SE & $\widehat\tau^2$(se)  \\
Hierarchical Meta-Analytic Approach&  -3.2 & 0.8 & 1.0 (2.3) \\
One-stage (full multivariate approach)&  -3.1 & 0.8 & 0.9 (2.3)  \\
       \hline
\hline
\end{tabular}
\end{table}

%% file: appendix.tex
\appendix{ Derivation of the covariance matrix for $\widehat{B}$}
The expression for the covariance between $\widehat{\theta}_{k}$ and $\widehat{\theta}_{l}$ 
is obtained based on a general formula for robust variance estimation. 
If we stack the score function in (\ref{eq-stageI-Sk-thetak}) we obtain
$S(\theta)=(S'_{1}(\theta_1), \ldots, S'_{K}(\theta_K))'$. 
Then given $B = (B_1, \ldots, B_K)'$ we note that
\begin{equation} \label{eq-sqrtJ-theta}
\sqrt{J} \, (\widehat{\theta}-\theta) \overset{d}{\sim} {\rm MVN}(0, \mathcal A^{-1}(\theta)  \, \mathcal B(\theta) \,  \mathcal A^{-1}(\theta))
\end{equation}
\noindent
as $J \rightarrow \infty$,
where 
$ \mathcal A(\theta)=E\{-\partial S(\theta)/\partial\theta'\}$ 
is a block diagonal $3K \times 3K$ matrix of the form
\begin{equation*}
{\cal A}(\theta) = 
\begin{bmatrix} 
  {\cal A}_{11}(\theta_{1}) &0                        & \dots  & 0 \\
  0                      &  {\cal A}_{22}(\theta_{2}) & \dots  & 0 \\
  \vdots                 &                         & \ddots & \vdots \\
  0                      &                         &        & {\cal A}_{KK}(\theta_{K})
\end{bmatrix}
\end{equation*}
\noindent
where the $k$th $3 \times 3$ diagonal sub-matrix is given by
\begin{equation*}
{\cal A}_{kk}(\theta_{k})
= E \{ -\partial S_k(\theta_k) / \partial \theta'_k \}
= E \left\{\sum_{j=1}^{J} R_{jk} \, X_{jk} \, X'_{jk} \right\} 
= J \, E\{X_{jk}\, X'_{jk} \mid R_{jk}=1\} \, P(R_{jk}=1) \,,
\end{equation*}
\noindent
$k=1,\ldots, K$.
If we let $\Omega_{kk} = P(X_{jk} \, X'_{jk} | R_{jk}=1)$ be a $3 \times 3$ matrix,
we can then write
\begin{equation} 
{\cal A}_{kk}(\theta_{k}) = J \, \Omega_{kk} \, P(R_{jk}=1) \,.
\end{equation}

Note that
$ \mathcal B(\theta)=E\{S(\theta) S'(\theta)\}$ is also a $3K \times 3K$ matrix. 
Under the assumption that the response data are missing at random
(i.e. $R_{jk} \perp Y_{jk} | X_{jk}$), 
the diagonal elements of ${\cal B}(\theta)$ are the covariance
matrices of the score functions for $\theta_k$, 
${\cal B}_{kk}(\theta) = {\rm cov}(S_k(\theta_k) \mid B)$,
$k = 1, \ldots, K$ where
\begin{equation*}
 {\cal B}_{kk}(\theta)
= E\left\{\sum_{j=1}^{J} R_{jk} \, S_{jk} (\theta_{k}) \, S'_{jk}(\theta_k)\right\}
= E \left\{\sum_{j=1}^{J} R_{jk} \, X_{jk}\, X'_{jk} \, {\rm var}(E_{jk}) \right\} \,,
\end{equation*}
\noindent
since the error terms are assumed independent of the covariates.
This can then be written as
\begin{equation}
{\cal B}_{kk}(\theta) = J \, \Omega_{kk} \, P(R_{jk}=1) \, \sigma^2_{k} \,, \quad k = 1, \ldots, K \,.
\end{equation}
\noindent
In a similar fashion we note that
\begin{align*}
{\cal B}_{kl}(\theta) = {\rm cov}(S_k(\theta_{k}), S_{l}(\theta_{l}) \mid B) 
& = E \left\{\sum_{j=1}^{J} R_{jk} \, R_{jl} \, E\{S_{jk} (\theta_{k}) S'_{jl} (\theta_{l}) \mid X_{jk}, X_{jl}, R_{jk}=R_{jl}=1\} \right\}\\
& = E \left\{\sum_{j=1}^{J}  R_{jk} \, R_{jl} \, X_{jk} \, X'_{jl} \,  {\rm cov}(E_{jk}, E_{jl}) \right\}\\
& = J \,  E\{X_{jk} \, X'_{jl} \mid R_{jk}=R_{jl}=1\} \, P(R_{jk}=R_{jl}=1)  \, \sigma_{kl} \\
& = J \,  \Omega_{kl} \, P(R_{jk}=R_{jl}=1)  \, \sigma_{kl}
\end{align*}
\noindent
where $\Omega_{kl}= E\{X_{jk} \, X'_{jl} | R_{jk}=R_{jl}=1\}$ is a $3 \times 3$ matrix.
If $X_{jk}=X_{jl}$ as in this setting, this becomes 
\begin{equation}
{\cal B}_{kl}(\theta) = {\rm cov}(S_{k}(\theta_{k}), S_{l}(\theta_{l}) \mid B)
= J \, \Omega_{kk} \, P(R_{jk}=R_{jl}=1) \, \sigma_{kl}  
\end{equation}
\noindent
since $\Omega_{kk} = \Omega_{kl}=\Omega$ for all $k \ne l$. 

If we wish to estimate the covariance of
$\sqrt{J} \, (\widehat{\theta}_k - \theta_k)$
and
$\sqrt{J} \, (\widehat{\theta}_l - \theta_l)$
given $B$ we note that this has the general form
\begin{equation*}
{\rm cov}(\sqrt{J} \, (\widehat{\theta}_k - \theta_k), \sqrt{J} \, (\widehat{\theta}_l - \theta_l) \mid B)
= {\cal A}_{kk}^{-1}(\theta) \, B_{kl}(\theta)  \, {\cal A}_{ll}^{-1}(\theta)  \,.
\end{equation*}
Inserting the derived expressions gives the $(k,l)$, $3 \times 3$ sub-matrix of the full covariance matrix in (\ref{eq-sqrtJ-theta}) as
\begin{equation} \label{eq-cov-sqrtJthetak-sqrtJthetal-given-B}
{\rm cov}(\sqrt{J} \, (\widehat{\theta}_{k}-\theta_{k}), \sqrt{J} \, (\widehat{\theta}_{l}-\theta_{l}) \mid B) 
=\dfrac{\sigma_{kl} \, \Omega^{-1} \, P(R_{jk}=R_{jl}=1)} {P(R_{jk}=1) \, P(R_{jl}=1)} \,.
\end{equation}

We estimate (\ref{eq-cov-sqrtJthetak-sqrtJthetal-given-B}) as follows. 
Since $X_{jk} = X_j$ is available for all individuals
we estimate $\Omega = \Omega_{kk} = \Omega_{kl}$ simply as
$\widehat{\Omega} = \sum_{j=1}^{J} (X_{jk}X'_{jk})/J$.
Moreover we estimate $P(R_{jk} = R_{jl} = 1)$ empirically as
$\widehat{P}(R_{jk}=R_{jl}=1)= n_{kl}/J$
where $n_{kl} = \sum_{j=1}^J R_{jk} R_{jl}$,
and likewise let $\widehat{P}(R_{jk}=1)= n_{k}/J$
where $n_k = \sum_{j=1}^J R_{jk}$, $k = 1, \ldots, K$.
Replacing unknown quantities with their estimates gives
\begin{equation} \label{eq-cov-sqrtJthetak-sqrtJthetal}
\widehat{\rm cov}(\sqrt{J} \, (\widehat{\theta}_{k}-\theta_{k}), \sqrt{J} \, (\widehat{\theta}_{l}-\theta_{l}) \mid B)
=\dfrac{\widehat{\sigma}_{kl}}{J^{-1} \, \sum_{j=1}^{J}(X_{jk} X'_{jl})}  \, \dfrac{J \, n_{kl}}{n_{k} n_{l}} 
\end{equation}
where 
$\widehat{\sigma}_{kl}$ is given by (\ref{eq-mle-sigmakl}).

Let $\mu(\beta) = \mathbb{I} \, \beta$ where $\mathbb{I}$ is a $K \times 1$ vector of ones and $\beta$ is a scalar.
We then let ${\rm cov}\{\sqrt{J} \, (\widehat{B}- \mu(\beta)) | B\}= \Gamma$ where 
$\Gamma$ is the covariance matrix for $\widehat{B}$ obtained by selecting the corresponding elements of (\ref{eq-cov-sqrtJthetak-sqrtJthetal-given-B})
related to the coefficients of the exposure variable
in the $K$ marginal least squares estimates.
We aim to use ${\rm cov}(\sqrt{J} \, (\widehat{B}- \mu(\beta)) | B)$ 
to combine the estimates across all responses,
but we note there is an additional component of variation in the estimators of the exposure effects
since the $B_k$ terms are themselves independent and identically distributed according to (\ref{eq-average-exposure-Bk}).
Thus while ${\rm cov}(\widehat{B}|B) = J^{-1} \, \Gamma$
where $\Gamma$ is a $K \times K$ matrix
with diagonal elements $\Gamma_{kk}$ and off-diagonal elements $\Gamma_{kl}$, $l = 1, \ldots, K$, $k = 1, \ldots, K$,
\begin{equation} \label{eq-varBk-hat}
{\rm var}(\widehat{B}_k) = J^{-1} \, \Gamma_{kk} + \phi \,, \quad k = 1, \ldots, K \,,
\end{equation}
\noindent
and since $B_k \perp B_l$,
\begin{equation} \label{eq-covBkBl-hat}
{\rm cov}(\widehat{B}_k, \widehat{B}_l) = J^{-1} \, \Gamma_{kl} \,, \quad k \neq l = 1, \ldots, K \,.
\end{equation}
\noindent
We denote the unconditional covariance matrix for $\widehat{B}$ as
${\rm cov}(\widehat{B}) = \Psi(\phi) = J^{-1} \, \Gamma + \Delta \, \phi$
where $\Psi_{kk}(\phi)$ is given by (\ref{eq-varBk-hat}), 
$\Psi_{kl}(\phi)$ is given by (\ref{eq-covBkBl-hat}),
and $\Delta$ is a $K \times K$ identity matrix.
Given an estimate of $\phi$,
we estimate this covariance matrix by 
\begin{equation*}
\widehat{\rm cov}( \widehat{B} ) = J^{-1} \, \widehat{\Gamma} + \Delta \, \widehat{\phi} = \widehat{\Psi}(\widehat{\phi}) \,.
\end{equation*}